\documentclass[twocolumn,final,showpacs,groupedaddress,prb]{revtex4}
\usepackage{amsmath}

\draft
\input{tcilatex}

\begin{document}

\title{SU(4) Spin-Orbital Two-Leg Ladder, Square and Triangle Lattices}
\author{Shun-Qing Shen}
\email{sshen@hkucc.hku.hk}
\affiliation{Department of Physics, The University of Hong Kong, Pokfulam Road, Hong
Kong, China}
\date{\today}

\begin{abstract}
Based on the generalized valence bond picture, a Schwinger boson mean field
theory is applied to the symmetric SU(4) spin-orbital systems. For a two-leg
SU(4) ladder, the ground state is a spin-orbital liquid with a finite energy
gap, in good agreement with recent numerical calculations. In
two-dimensional square and triangle lattices, the SU(4) Schwinger bosons
condense at $(\pi/2,\pi/2)$ and $(\pi/3,\pi/3)$, respectively. Spin,
orbital, and coupled spin-orbital static susceptibilities become singular at
the wave vectors, twice of which the bose condensation arises at. It is also
demonstrated that there are spin, orbital, and coupled spin-orbital
long-range orderings in the ground state.
\end{abstract}

\maketitle

Recently the properties of spin systems with orbital degeneracy are
attracting a lot of attentions.\cite{Tokura01} Several spin-orbital models
are proposed in various kinds of materials, such as C$_{60},$\cite{Arovas95}
NiLiO$_{3},$ \cite{Li98} Na$_{2}$Ti$_{2}$Sb$_{2}$O, \cite{Pati98} and LaMnO$%
_{3}$.\cite{Ishihara97} Interplay of spin and orbital degrees of freedom
produces not only new magnetic structure phases but also novel quantum
ordered and disordered states such as orbital density wave and spin-orbital
liquids. A simplified and symmetric model for these systems is \cite%
{Kugel73,Castellani78} 
\begin{equation}
H=\frac{1}{2}\sum_{i,\mathbf{\delta }}J_{\mathbf{\delta }}\left( 2\mathbf{S}%
_{i}\cdot \mathbf{S}_{i+\mathbf{\delta }}+1/2\right) \left( 2\mathbf{T}%
_{i}\cdot \mathbf{T}_{i+\mathbf{\delta }}+1/2\right) ,  \label{su4}
\end{equation}%
where the operators $\mathbf{S}$ and $\mathbf{T}$ are SU(2) Pauli matrices
for the spin and orbital degrees of freedom, respectively. The vector $%
\delta $ points to the nearest neighboring sites. It is already known that
the model possesses SU(4) symmetry, and can be derived from a quarter-filled
electronic model with two-fold orbital degeneracy by ignoring the Hund's
rule couplings and taking the large on-site Coulomb interaction. High
symmetry in this spin-orbital model means strong correlations between spins
and orbitals. In an SU(4) symmetric state, the correlation functions for
fifteen generators of the SU(4) Lie group are isotropic. The spin, orbital,
and coupled spin-orbital degrees of freedom must be treated on equal footing.%
\cite{Li98,Zhang01} Over last a few years this model was studied
extensively. In one dimension (1D), it is fairly understood analytically and
numerically.\cite{Sutherland75,Azaria99,itoi,Frischmuth99,Zheng01} The 1D
model can be solved by means of Bethe ansatz, and its ground state is
described by a gapless spin liquid, similar to the SU(2) Heisenberg model.
In two dimension (2D), it is relatively less understood. Li, \textit{et al.} 
\cite{Li98} first argued that an SU(4) singlet plaquette state contains at
least four sites, and a collection of such SU(4) singlets may lead to a spin
liquid state. This picture is realized very well in two-leg ladder model, 
\cite{Bossche01} and some solvable models.\cite{Shen01a} So far it is not
clear whether such an idea can be realized in 2D. Except numerical
diagonalization on small clusters \cite{Bossche00} and series expansion, 
\cite{Zasinas01} there has been no solid results as whether the ground state
is long-range ordered or a spin-orbital liquid.

In this paper, the properties of the generalized valence bond state
consisting of the SU(4) singlets are discussed in detail for the model (Eq.(%
\ref{su4})) in a SU(4) Schwinger boson mean field theory. The ground state
of the system can be either ordered or disordered, depending on the
dimensionality and lattice topological structure. In a two-leg ladder
system, we find that the ground state is a spin liquid state with a finite
energy gap, which is in good agreement with recent numerical calculations by
van den Bossche.\cite{Bossche01} In square and triangle lattices, the
Schwinger bosons condense at zero temperatures, \textit{i.e}., the
Bose-Einstein condensation (BEC) occurs and is identified as the indication
of LROs in the ground state. The relation between BEC and LROs is
illustrated explicitly in spin, orbital, and spin-orbital static transverse
susceptibilities, which become singular at the wave vectors ($\pi /2,\pi /2$%
) and ($\pi /3,\pi /3$) for the respective square and triangle lattices,
leading to finite staggered magnetizations for the spin, orbital, and
coupled spin-orbital densities in the thermodynamic limit. Thus, three
Goldstone modes are found. According to the calculated susceptibilities, the
spin, orbital, and spin-orbital LROs may coexist, but the direction of
spontaneous symmetry breaking will determine the properties of the ground
state.

In general, for the spin-1/2 system with double orbital degeneracy, there
are four local states on each site $i$ according to the eigenvalues of $%
\mathbf{S}_{i}^{z}$ and $\tau _{i}^{z}$: $\left\vert 1\right\rangle
=\left\vert +1/2,+1/2\right\rangle $, $\left\vert 2\right\rangle =\left\vert
-1/2,+1/2\right\rangle $, $\left\vert 3\right\rangle =\left\vert
+1/2,-1/2\right\rangle $, $\left\vert 4\right\rangle =\left\vert
-1/2,-1/2\right\rangle $. Four Schwinger bosons can be introduced to
describe these four states: $\left\vert \mu \right\rangle =a_{\mu }^{\dagger
}\left\vert 0\right\rangle $ where $\left\vert 0\right\rangle $ is the
vacuum states and $\mu =1,2,3,4$. There has to be imposed a local
constraint, $\sum_{\mu =1}^{4}a_{i\mu }^{\dagger }a_{i\mu }=1$ on each
lattice site. The permutation operator $P_{ij}=\left( 2\mathbf{S}_{i}\cdot 
\mathbf{S}_{j}+\frac{1}{2}\right) \left( 2\mathbf{T}_{i}\cdot \mathbf{T}_{j}+%
\frac{1}{2}\right) $ is to exchange the two states on the sites $i$ and $j$, 
$P_{ij}\left\vert i\mu ,j\nu \right\rangle =\left\vert i\nu ,j\mu
\right\rangle $. Moreover, $P_{ij}$ can be expressed in terms of the four
hard-core bosons as $P_{ij}=\sum_{\mu ,\nu }a_{i\mu }^{\dagger }a_{i\nu
}a_{j\nu }^{\dagger }a_{j\mu }$. An SU(4) singlet is defined by $%
SU_{4}(i,j,k,l)=\sum_{\mu ,\nu ,\gamma ,\delta }\Gamma _{\mu ,\nu ,\gamma
,\delta }$ $a_{i\mu }^{\dagger }a_{j\nu }^{\dagger }a_{k\gamma }^{\dagger
}a_{l\delta }^{\dagger }\left\vert 0\right\rangle $, where $\Gamma _{\mu
,\nu ,\gamma ,\delta }$ is an antisymmetric tensor. When the model
Hamiltonian Eq.(\ref{su4}) has only four lattice sites, an SU(4) singlet is
always the lowest energy state for $J_{\mathbf{\delta }}\geq 0$. According
to the group theory, the SU(4) symmetric state for a lattice with $4n$ sites
($n$ is integer) can be regarded as a linear combination of all states
consisting of $n$ SU(4) singlets.\cite{Shen01a} This is a generalization of
the Anderson's resonating valence bond (VB) state \cite{Anderson73} from the
spin SU(2) system to the SU(4) system. As is well-known in the Heisenberg
model, a short-range VB state may describe a spin liquid state with a finite
energy gap,\cite{Affleck87} and a long-range VB state may possess
antiferromagnetic LRO.\cite{Liang88} A Schwinger boson mean field theory
based on the short-range VB state was proposed for the spin SU(2) systems by
Auerbach and Arovas,\cite{Auerbach88} which successfully describes either
ordered or disordered quantum states. Very recently, the theory was applied
to the spin-orbital systems by the present author and his collaborator.\cite%
{Zhang01}

To realize the generalized VB state in the SU(4) spin-orbital system, the
model Hamiltonian is rewritten as 
\begin{eqnarray}
H &=&-\frac{1}{4}\sum_{i,\mathbf{\delta },\mu ,\nu }J_{\mathbf{\delta }}%
\text{ }A_{i,i+\mathbf{\delta };\mu ,\nu }^{\dagger }\text{ }A_{i,i+\mathbf{%
\delta };\mu ,\nu }  \notag \\
&&+\sum_{i}\lambda _{i}(\sum_{\mu =1}^{4}a_{i,\mu }^{\dagger }a_{i,\mu }-1)+%
\frac{1}{2}N_{\Lambda }\sum_{\mathbf{\delta }}J_{\mathbf{\delta }},
\end{eqnarray}%
where $A_{i,j;\mu ,\nu }=a_{i\mu }a_{j\nu }-a_{i\nu }a_{j\mu }$ and $%
N_{\Lambda }$ is the total number of lattice sites. Antisymmetric operators $%
A_{i,j;\mu ,\nu }$ are introduced for the purpose of the mean field
calculations. The following theory is limited to the case $J_{\mathbf{\delta 
}}\geq 0$. The local Lagrangian multiplier is also used to impose the local
constraint for the hard-core bosons on average. In the mean field
approximation we will take all $\lambda _{i}=\lambda $. The thermodynamic
averages of the operators $A_{ij,\mu \nu }$ are defined as the VB order
parameters, $\left\langle A_{i,i+\mathbf{\delta };\mu ,\nu }\right\rangle
\equiv -2i\Delta _{\mu ,\nu }(\mathbf{\delta })$, which are odd functions
with respect to either the indices $\mu $, $\nu $ or the vector direction $%
\mathbf{\delta }$. In the momentum space, we define 
\begin{equation*}
\gamma _{\mu ,\nu }(\mathbf{k})\equiv 2i\sum_{\mathbf{\delta }}J_{\mathbf{%
\delta }}\Delta _{\mu \nu }(\mathbf{\delta })\exp (i\mathbf{k}\cdot \mathbf{%
\delta }).
\end{equation*}%
We define an eight-component spinor 
\begin{equation*}
\Phi _{\mathbf{k}}^{\dagger }=(a_{\mathbf{k}1}^{\dagger },a_{\mathbf{k}%
2}^{\dagger },a_{\mathbf{k}3}^{\dagger },a_{\mathbf{k}4}^{\dagger },a_{-%
\mathbf{k}1},a_{-\mathbf{k}2},a_{-\mathbf{k}3},a_{-\mathbf{k}4}).
\end{equation*}%
By utilizing the Pauli matrices $\sigma _{\alpha }$ ($\alpha =x,y,z$), the
decoupled mean field Hamiltonian is thus written in a compact matrix form, 
\begin{equation}
H=\frac{1}{2}\sum_{\mathbf{k}}\Phi _{\mathbf{k}}^{\dagger }(\lambda -i\sigma
_{y}\otimes \mathbf{B}(\mathbf{k}))\Phi _{\mathbf{k}}+\mathcal{E}_{0}
\end{equation}%
where $\mathcal{E}_{0}/N_{\Lambda }=\sum_{\mathbf{\delta }}J_{\mathbf{\delta 
}}\Delta _{\mu \nu }^{2}(\mathbf{\delta })-3\lambda +\frac{1}{2}\sum_{%
\mathbf{\delta }}J_{\mathbf{\delta }};$%
\begin{equation*}
\mathbf{B}(\mathbf{k})=\left( 
\begin{array}{cccc}
0 & \gamma _{12}(\mathbf{k}) & \gamma _{13}(\mathbf{k}) & \gamma _{14}(%
\mathbf{k}) \\ 
-\gamma _{12}(\mathbf{k}) & 0 & \gamma _{23}(\mathbf{k}) & \gamma _{24}(%
\mathbf{k}) \\ 
-\gamma _{13}(\mathbf{k}) & -\gamma _{23}(\mathbf{k}) & 0 & \gamma _{34}(%
\mathbf{k}) \\ 
-\gamma _{14}(\mathbf{k}) & -\gamma _{24}(\mathbf{k}) & -\gamma _{34}(%
\mathbf{k}) & 0%
\end{array}%
\right) ,
\end{equation*}%
where $N_{\Lambda }$ is the number of lattice sites. Considering the
symmetry in the Hamiltonian,\cite{degeneracy} there exists two sets of
solutions: (I). $\gamma _{12}(\mathbf{k})=\gamma _{34}(\mathbf{k}),$ $\gamma
_{13}(\mathbf{k})=-\gamma _{24}(\mathbf{k}),$ $\gamma _{14}(\mathbf{k}%
)=\gamma _{23}(\mathbf{k});$ (II). $\gamma _{12}(\mathbf{k})=-\gamma _{34}(%
\mathbf{k}),$ $\gamma _{13}(\mathbf{k})=\gamma _{24}(\mathbf{k}),$ $\gamma
_{14}(\mathbf{k})=-\gamma _{23}(\mathbf{k}).$ The physical reason is that an
SU(4) singlet plaquette state $SU_{4}(i,j,k,l)$ contains four creation
operators with different sites and indices, and $A_{ij,\mu \nu }^{\dagger }$
contains only two sites and two indices. To form such an SU(4) singlet
plaquette state, the four creation operators on different sites should have
different indices. The relations in the solution (I) and (II) reflect these
properties. In each SU(4) singlet plaquette it consists of two
configurations. Each configuration consists of two spin singets and two
orbital singlets. The two configurations are degenerated, but not
orthogonal. The double degeneracy of the solutions may be related to the
properties of SU(4) singlet plaquettes. So these relations will help us to
construct the wave function of the generalized SU(4) VB state. We first
focus on the solution (I), and will discuss the results of the solution
(II). The single-particle Green function is given by 
\begin{eqnarray}
&&G(\mathbf{k},i\omega _{n})\equiv \left[ i\omega _{n}\Omega _{1}-\lambda
+i\sigma _{y}\otimes \mathbf{B}(\mathbf{k})\right] ^{-1}  \notag \\
&=&\frac{i\omega _{n}\Omega _{1}+\lambda +\gamma _{12}(\mathbf{k})\Omega
_{2}+\gamma _{13}(\mathbf{k})\Omega _{3}+\gamma _{13}(\mathbf{k})\Omega _{4}%
}{(i\omega _{n})^{2}-\omega (\mathbf{k})^{2}},  \label{Green}
\end{eqnarray}%
where $\Omega _{1}=\sigma _{z}\otimes \sigma _{0}\otimes \sigma _{0}$, $%
\Omega _{2}=\sigma _{y}\otimes \sigma _{0}\otimes \sigma _{y}$, $\Omega
_{3}=\sigma _{y}\otimes \sigma _{y}\otimes \sigma _{z}$, and $\Omega
_{4}=\sigma _{y}\otimes \sigma _{y}\otimes \sigma _{x}$. There is only one
four fold degenerate quasiparticle spectrum: 
\begin{equation}
\omega (\mathbf{k})=\sqrt{\lambda ^{2}-\ [\gamma _{12}^{2}(\mathbf{k}%
)+\gamma _{13}^{2}(\mathbf{k})+\gamma _{14}^{2}(\mathbf{k})]},
\end{equation}%
from which the free energy for the system is evaluated 
\begin{equation}
F=\frac{4}{\beta }\sum_{\mathbf{k}}\ln [1-\exp [-\beta \omega (\mathbf{k}%
)]+2\sum_{\mathbf{k}}\omega (\mathbf{k})+\mathcal{E}_{0}.  \label{free}
\end{equation}%
The saddle point equations are thus derived by minimizing the free energy
with respect to the mean field variables $\lambda $ and $\Delta _{\mu \nu }(%
\mathbf{\delta })$. We can deduce the VB order parameters according to the
symmetry of $\Delta _{\mu \nu }(\mathbf{\delta })$ and of lattice.

Now we apply the general formalism to the SU(4) model on several lattices.
We first study the two-leg ladder model. Recent numerical study has shown
that its ground state is a spin-orbital liquid with a finite energy gap,\cite%
{Bossche01} which can be regarded as a realization of SU(4) plaquette state,
or a short-range generalized VB state. Here the isotropic case is considered 
$J_{\parallel }=J_{\perp }=J$ so that we introduce two sets of VB order
parameters: $\Delta _{\mu \nu }(x)$ along the ladder and $\Delta _{\mu \nu
}(y)$ along the rungs. Special attention should be paid for the direction
along the rung. The momentum along the rung has two discrete values. The
spectra are given by 
\begin{equation}
\omega _{\pm }(\mathbf{k})=\sqrt{\lambda ^{2}-16J^{2}\Delta _{\parallel
}^{2}(\sin \mathbf{k}\pm \eta )^{2}}
\end{equation}%
where $\Delta _{\parallel }^{2}=\Delta _{12}^{2}(x)+\Delta
_{13}^{2}(x)+\Delta _{14}^{2}(x)$ and $\eta =\Delta _{\mu \nu }(y)/2\Delta
_{\mu \nu }(x)$ is determined by the mean field equations \cite{Note}. Two
branches of spectra have a relation: $\omega _{+}(\mathbf{k})=\omega _{-}(-%
\mathbf{k})$. The saddle point equations are determined by minimizing the
free energy with respect to the mean field variables

\begin{subequations}
\begin{eqnarray}
\int \frac{d\mathbf{k}}{2\pi }\frac{\lambda }{\omega _{+}(\mathbf{k})}\ %
\left[ 2n_{B}(\omega _{+}(\mathbf{k}))+1\right] &=&\frac{3}{2}, \\
\ \int \frac{d\mathbf{k}}{2\pi }\frac{\ (\sin \mathbf{k}+\eta )^{2}}{\omega
_{+}(\mathbf{k})}\ \ \left[ 2n_{B}(\omega _{+}(\mathbf{k}))+1\right] &=&%
\frac{1+2\eta ^{2}}{2J}, \\
\int \frac{d\mathbf{k}}{2\pi }\frac{\sin \mathbf{k}+\eta }{\omega _{+}(%
\mathbf{k})}\ \ \left[ 2n_{B}(\omega _{+}(\mathbf{k}))+1\right] &=&\frac{%
\eta }{J}.
\end{eqnarray}%
where $n_{B}(x)$ is the distribution function for bosons: $n_{B}(x)=1/[\exp
[\beta x]-1]$ and $\beta =1/k_{B}T$. At $T=0,$ $n_{B}(\omega _{+}(\mathbf{k}%
))=0$ when $\omega _{+}(\mathbf{k})\neq 0.$ The numerical calculations give
rise to $\lambda =2.26574J$, $\Delta _{\parallel }=0.35339$, and $\eta
=0.55146$. The minima of spectra is at $\mathbf{k}^{\ast }=\pi /2$ for $%
\omega _{+}(\mathbf{k})$, and at $-\pi /2$ for $\omega _{-}(\mathbf{k})$: $%
\min (\omega _{\pm }(\mathbf{k}))$ $=\sqrt{\lambda ^{2}-16J^{2}\Delta
_{\parallel }^{2}(1+\eta )^{2}}$. From the dynamic susceptibilities of spin,
orbital, and spin-orbital operators, we find that there is a finite energy
gap: $\Delta _{GAP}=2\min (\omega _{\pm }(\mathbf{k}))=1.138J$. For the
second set of solution (II), it also produces the same numerical results,
degenerated with the first set of solutions. van den Bossche \textit{et. al}%
. \cite{Bossche01} studied this SU(4) ladder model up to $16$ sites by an
exact diagonalization method, and a finite energy gap has been found in a
singlet-multiplet excitation: $\Delta =1.09J$. The two values are in
excellent agreement. A local minimum at $\mathbf{k}=\pi /2$ in the
quasiparticle dispersion was also observed. Moreover, the ground state has
also been found to have a two-fold degeneracy in the thermodynamic limit,
consistent with the two solutions of ours. All these facts can be regarded
as strong support for our present theory. It is worth noting that the theory
may fail for one-dimensional chain to predict a small energy gap due to the
ignorance of the topological terms or \textquotedblleft Pontryagin
index\textquotedblright , which can destroy the Haldane's gap. The same
problem was encountered in the SU(2) theory, and was discussed extensively
in one-dimensional spin 1/2 systems.\cite{Auerbach88}

Next we come to study 2D lattices. Let us consider a square lattice first.
In this case, we still assume isotropic couplings $J_{x}=J_{y}$. The
spectrum for the bosonic quasi-particles can be written as\cite{Rotation} 
\end{subequations}
\begin{equation}
\omega (\mathbf{k})=\sqrt{\lambda ^{2}-16J^{2}\Delta ^{2}(\sin \mathbf{k}%
_{x}+\sin \mathbf{k}_{y})^{2}},
\end{equation}%
with $\Delta ^{2}=\Delta _{12}^{2}+\Delta _{13}^{2}+\Delta _{14}^{2}$. The
minimum of the energy spectra occurs at $\mathbf{k}^{\ast }=(\pi /2,\pi /2%
\dot{)}$. The saddle point equations are given by, 
\begin{eqnarray}
\int \frac{d\mathbf{k}}{(2\pi )^{2}}\frac{\lambda }{\omega (\mathbf{k})}\ %
\left[ 2n_{B}(\omega (\mathbf{k}))+1\right]  &=&\frac{3}{2},  \label{f1} \\
\ \int \frac{d\mathbf{k}}{(2\pi )^{2}}\frac{(\sin \mathbf{k}_{x}+\sin 
\mathbf{k}_{y})^{2}}{\omega (\mathbf{k})}\ \ \left[ 2n_{B}(\omega (\mathbf{k}%
))+1\right]  &=&\frac{z}{4J},  \label{f2}
\end{eqnarray}%
where the coordinate number for square lattice is $z=4$. In the present
theory the number of bosons in the diagonalized Hamiltonian is not equal to
the number of the hard-core bosons since the Bogoliubov transformation
changes the number of bosons. The quasi-particle number is determined by
solutions of the saddle point equations self-consistently. At $T=0$ the
bosons may condense, \textit{i.e.}, the BEC occurs. The minima of the boson
spectra $\omega (\mathbf{k})$ are at $\mathbf{k}=\pm \mathbf{k}^{\ast }$.
Since the distribution function becomes singular at $\omega (\mathbf{k}%
^{\ast })=0$ in Eq.(\ref{f1}), we have to introduce a finite quantity $%
b_{0}=2\left[ n_{B}(\omega (\mathbf{k}^{\ast }))+n_{B}(\omega (-\mathbf{k}%
^{\ast }))\right] /\left[ N_{\Lambda }\omega (\mathbf{k}^{\ast })\right] $
such that the saddle point equations have physical solutions. The non-zero $%
b_{0}$ may be used as the order parameters for long-range correlations for
the spin, orbital, and coupled spin-orbital densities. The saddle point
equations are solved numerically. When $T>0$, $b_{0}=0$. At $T=0$, we have $%
\lambda =8J\Delta =3.05905J$ and $b_{0}=0.1068$. The non-zero $b_{0}$
indicates that the BEC\ occurs on a square lattice.

Another typical 2D lattice is the triangle lattice. For example, NiLiO$_{3}$
has a 2D triangle lattice structure, and was modeled as a spin-orbital
system \cite{Li98,Reynaud01}. Usually the quantum frustration is anticipated
to make quantum fluctuations more stronger. Topologically, we can distort a
triangle lattice into a square one by introducing a finite diagonal coupling 
$J_{x+y}$, and take another diagonal coupling $J_{x-y}=0$. In the present
theory we have three sets of order parameters: $\Delta _{\mu \nu }(\mathbf{x}%
)$, $\Delta _{\mu \nu }(\mathbf{y})$, and $\Delta _{\mu \nu }\mathbf{(x+y})$%
. Focus on the isotropic case $J_{x}=J_{y}=J_{x+y}$, then the VB order
parameters have a relation: $\Delta _{\mu \nu }(\mathbf{x})=\Delta _{\mu \nu
}(\mathbf{y})=\Delta _{\mu \nu }(\mathbf{x+y})=\Delta _{\mu \nu }$. The
spectra for the bosonic quasi-particles is thus given by 
\begin{equation}
\omega (\mathbf{k})=\sqrt{\lambda ^{2}-16J^{2}\Delta ^{2}(\sin \mathbf{k}%
_{x}+\sin \mathbf{k}_{y}+\sin (\mathbf{k}_{x}+\mathbf{k}_{y}))^{2}},
\end{equation}%
with $\Delta ^{2}=\Delta _{12}^{2}+\Delta _{13}^{2}+\Delta _{14}^{2}$. The
minimum of the spectra occurs at $\mathbf{k}^{\ast }=\pm (\pi /3,\pi /3)$,
which can be shifted away if the couplings are anisotropic. A similar set of
saddle point equations are obtained by minimizing the free energy in Eq. (%
\ref{free}). Numerically solving the self-consistent equation at $T=0$ gives
rise to $\lambda =6\sqrt{3}J\Delta =3.57878J$ and $b_{0}=0.155345$. Since $%
\omega (\mathbf{k}^{\ast })=0$, the BEC also appears on an isotropic
triangle lattice. The coupling $J_{x+y}$ does not enhance the quantum
frustration to suppress the BEC completely. The role of $J_{x+y}$ is to
force the minimal point from $\pm (\pi /2,\pi /2\dot{)}$ for $J_{x+y}=0$ to $%
\pm (\pi /3,\pi /3)$ for $J_{x+y}=J_{x}=J_{y}$. For a more detailed
calculations, it is found that the wave vector $\mathbf{k}^{\ast }$ changes
continuously as a function of $J_{x+y}/J_{x}$.

In the usual Schwinger boson mean field theory the BEC is identified as the
long-range correlations between the SU(2) spins. To establish the relation
between the BEC and the long-range orders in the present SU(4) ground state,
we calculated the static susceptibilities \cite{Schrieffer84} 
\begin{eqnarray}
&&\chi _{X}(\mathbf{q})=-\frac{1}{16N_{\Lambda }}\sum_{\mathbf{k}}  \notag \\
&&\times \text{Tr}\left\{ \Omega _{X}G(\mathbf{k}+\mathbf{q},\tau
=0^{-})\Omega _{X}G(\mathbf{k},\tau =0^{+}\right\} ,
\end{eqnarray}%
where $X=S$ for spin $\mathbf{S}_{i}^{z}$, $X=T$ for orbital $\mathbf{T}%
_{i}^{z}$, and $X=ST$ for the coupled spin-orbital $2\mathbf{S}_{i}^{z}%
\mathbf{T}_{i}^{z}$. These three operators can be expressed in terms of
spinors, $\Phi _{i}^{\dagger }\Omega _{X}\Phi _{i}/4,$ where $\Omega
_{S}=\sigma _{0}\otimes \sigma _{0}\otimes \sigma _{z}$, $\Omega _{T}=\sigma
_{0}\otimes \sigma _{z}\otimes \sigma _{0}$, and $\Omega _{ST}=\sigma
_{0}\otimes \sigma _{z}\otimes \sigma _{z}$. From the single particle Green
function Eq.(\ref{Green}), we can calculate the static susceptibilities, 
\begin{subequations}
\begin{eqnarray}
\chi _{S}(\mathbf{Q})/N_{\Lambda } &\approx &\frac{1}{4}b_{0}^{2}\gamma
_{13}^{2}(\mathbf{q}=2\mathbf{k}^{\ast }), \\
\chi _{T}(\mathbf{Q})/N_{\Lambda } &\approx &\frac{1}{4}b_{0}^{2}\gamma
_{12}^{2}(\mathbf{q}=2\mathbf{k}^{\ast }), \\
\chi _{ST}(\mathbf{Q})/N_{\Lambda } &\approx &\frac{1}{4}b_{0}^{2}\gamma
_{14}^{2}(\mathbf{q}=2\mathbf{k}^{\ast }),
\end{eqnarray}%
which become singular when $\mathbf{q}=\mathbf{Q}=2\mathbf{k}^{\ast }$ and
higher order terms can be ignored. The correlation functions are
proportional to the number of lattice sites once $b_{0}\neq 0$. These
properties are characterized by long-range correlations at the wave vector $%
\mathbf{Q}$. In the thermodynamic limit, the corresponding magnetizations
become $m_{X}=\sqrt{\chi _{X}(Q)/N_{\Lambda }}$, which depend on the values
of VB order parameters, $\Delta _{12}^{2}$, $\Delta _{13}^{2}$, and $\Delta
_{14}^{2}$, respectively. The long-range order is thus at $\mathbf{Q}=(\pi
,\pi )$ for a square lattice ($(\pi ,\pi )$ and $(-\pi ,-\pi )$ are the same
vector), and $\mathbf{Q}=(2\pi /3,2\pi /3)$ or $(-2\pi /3,-2\pi /3)$ for a
triangle lattice. The two vectors $\pm $ $k^{\ast }$ correspond to one state
for a square lattice, but two equivalent states for a triangle lattice. From
the second set of mean field solutions, we find that the above relations
remain if we make a permutation between $\mathbf{S}$ and $\mathbf{T}$.
Therefore these two solutions (I) and (II) are degenerated. We believe that
the double degeneracy of the ground states observed in our theory is not a
result of the mean field theory, and may have a deep physical origin. The
singularity in static susceptibilities also reflects the fact that the
collective modes are gapless Goldstone modes. The SU(4) system may have at
most three Goldstone mode when the symmetry is broken spontaneously.

When $\Delta _{12}^{2}=\Delta _{13}^{2}=\Delta _{14}^{2}=\Delta ^{2}/3$, we
have $\chi _{S}=\chi _{T}=\chi _{ST}$. These relations are in agreement with
the isotropic correlations for spin, orbital and coupled spin-orbital in our
SU(4) invariant ground state. In such a mean field theory, the operator $%
\mathbf{S}^{z}$, $\mathbf{T}^{z}$, and $2\mathbf{S}^{z}\mathbf{T}^{z}$ play
an equal role. When spin and orbital density operators have long-range
correlations, and the coupled spin-orbital density operators also have
long-range correlations with the same wave vectors. An interesting
observation is that the ground state energy depends only on the parameter $%
\Delta ^{2}.$\cite{Shen96} It may contain some new states, which is
determined by the direction of spontaneous symmetry breaking.\cite%
{Anderson84} Actually, we have the freedom to choose the direction of the
spontaneous symmetry breaking in the thermodynamic limit. For example, an
infinitesimal external staggered magnetic field along the spin z-direction
may induce a ground state with $\Delta _{12}^{2}=\Delta ^{2}$ and $\Delta
_{13}^{2}=\Delta _{14}^{2}=0$, where a magnetic long-range order appears
only. An infinitesimal John-Teller distortion may induce a ground state with 
$\Delta _{13}^{2}=\Delta ^{2}$ and $\Delta _{12}^{2}=\Delta _{14}^{2}=0$,
where an orbital long-range order shows up. It is also possible for two or
more types of long-range orders to coexist in the a single ground state.

In conclusion, based on a generalized valence bond state picture, a
Schwinger boson mean field theory is developed for the symmetric SU(4)
spin-orbital systems, showing that the ground state for a two-leg ladder
model is a spin-orbital liquid with a finite energy gap in low energy
excitations, and the ground states for square and triangle lattices possess
spin, orbital, and coupled spin-orbital long-range orderings.

I would like to thank G. M. Zhang for his helpful discussions. This work was
supported by a RGC grant of Hong Kong, and a CRCG\ grant of The University
of Hong Kong.

\end{subequations}


\begin{thebibliography}{99}
\bibitem{Note} $\eta $ is independent of $\mu \nu $ which can be derived
from the mean field equations if we list all equations for $\Delta _{\mu \nu
}(\mathbf{\delta })$.
\bibitem{Rotation} The spectra in Eqs. (9) and (11) have no rotation
symmtries of square and triangle lattices. Under the rotation symmetry we
have several other solutions which give the same physical consequences. It
is worthy noting that the dispersion relations for Schwinger bosons are not
observables and physical observables are related to the spin operators
explicitly.
\end{thebibliography}
\end{document}